\documentclass[12pt]{article}
\usepackage{amsmath}
\usepackage{amssymb}
\usepackage{amsfonts, amsthm}
\begin{document}
\begin{center}{\large\bf {Energy Spectrum of Quasi-Geostrophic Turbulence}
}\end{center} 
\begin{center}{\large\bf Peter Constantin}\end{center}
\begin{center}{Department of Mathematics}\end{center}
\begin{center}{The University of Chicago}\end{center}
\begin{center}{7/16/02}\end{center}
\vspace{1cm}

\noindent{\bf Abstract.}
We consider the energy spectrum of a quasi-geostrophic model of forced,
rotating turbulent flow. We provide a rigorous a priori bound $E(k)\le Ck^{-2}$
valid for wave numbers that are smaller than a wave number associated to the
forcing injection scale. This upper bound separates this spectrum from
the Kolmogorov-Kraichnan $k^{-\frac{5}{3}}$ energy spectrum 
that is expected in a two-dimensional Navier-Stokes inverse cascade. 
Our bound provides theoretical support for the $k^{-2}$ spectrum observed in 
recent experiments.

\vspace{1cm}
 
The typical time scales associated with atmospheric flow over long
distances are much bigger than the time scales associated with Earth's
rotation. This low Rossby number situation is characterized by a 
relative suppression of momentum transfer across vertical scales,
and the organization of the flow in quasi-two dimensional strata. 
Cyclonic and anti-cyclonic vortical motion ensues in these layers, with 
dynamics in which strong, interacting vortices of many sizes are born, grow, 
and dissipate, over time scales that are long compared to the rotation 
time scale. The precise mathematical way of describing 
such a quasi-two dimensional picture is yet unclear. 
Energy spectra are some of the most robust quantitative indicators that 
one can use in order to distinguish between different classes of 
models.  If a strictly two dimensional Navier-Stokes framework is 
adopted for rotating turbulence then the predicted energy spectra are a $k^{-3}$ direct enstrophy cascade (at wave numbers larger than the wave number of 
the forces stirring the fluid) and a Kolmogorov-Kraichnan 
$k^{-{\frac{5}{3}}}$ inverse energy cascade spectrum at wave numbers that 
are smaller than the forcing wave numbers \cite{Fr}.  Recent experiments \cite{Sw} of rotating fluids find a different inverse energy cascade power spectrum: 
$E\sim k^{-2}$. This spectrum implies a steeper inverse energy cascade than
the one predicted by a strictly two-dimensional Kolmogorov-Kraichnan 
spectrum. The $k^{-2}$ spectrum was observed for wave 
numbers of $10^{-1} - 10^{0}\,$ cm$^{-1}$. The experimental data showed the 
steeper $k^{-2}$ spectrum clearly separated from a $k^{-\frac{5}{3}}$ 
spectrum that was fit to agree with $E(k_0)$ at the largest scale $k_0\sim 10^{-1}\,$ cm$^{-1}$.

The purpose of this letter is to describe a
rigorous upper bound $E(k) \le C k^{-2}$ valid in the inverse cascade 
region, in a quasi-geostrophic regime.

The most important feature of strongly rotating fluids is the 
geostrophic balance between the Coriolis force and pressure gradients.
This balance, valid only in a first approximation, imposes
a two-dimensional time independent solution. The departure
from this balance, to lowest order, has non-trivial dynamics and is 
described by quasi-geostrophic equations \cite{PePi}, which are quasi-two 
dimensional equations asserting the conservation of potential vorticity $q$ 
subject to dynamical boundary conditions.  The simplest of these
can be written as
\begin{equation}
\partial_t\theta + {\mathbf v}\cdot\nabla \theta + w_E\Lambda\theta = f
\label{qg}
\end{equation}
The two-dimensional velocity  ${\mathbf v}$ is incompressible, $\nabla
\cdot{\mathbf v}= 0$. The dissipative term $w_E\Lambda\theta$ has
a coefficient $w_E>0$ that comes from Ekman pumping at the boundary. 
This coefficient has units of velocity and, for the situations we 
consider, where a non-trivial vertical flux is imposed, this coefficient is 
not vanishingly small. (In the experiment (\cite{Sw}) the vertical velocities at the boundary are close in magnitude to the maximal measured velocities, and are in ranges of about 20-30 cm/s; the value for $w_E$ is expected
to match the same order of magnitude). 
The operator $\Lambda $ can be described in Fourier representation as 
multiplication by the magnitude $k$ of the wave number ${\mathbf k} = (k_x, k_y)$, that is: 
\begin{equation}
{\widehat{\Lambda\theta}}({\mathbf k},t) = k{\widehat{\theta}}({\mathbf{k}},t).
\label{lambda}
\end{equation}
The velocity is related to $\theta$ by  
\begin{equation}
{\widehat{{\mathbf v}}}({\mathbf{k}}, t) = (-k_y,k_x)\frac{\sqrt{-1}}{k}\widehat{\theta}({\mathbf k}, t)\label{actsc}
\end{equation}
Thus, for each wave number 
\begin{equation}
|{\widehat{\theta}}({\mathbf k},t)| = | {\widehat{{\mathbf v}}}({\mathbf{k}}, t)|. \label{tveq}
\end{equation}
The symbol $f$ refers to the forcing term. 
This active scalar surface quasi-geostrophic equation has been studied both 
analytically and numerically \cite{CW1}. 

We will analyze the energy spectrum using the Littlewood-Paley decomposition
\cite{MT}.  The Littlewood-Paley decomposition is not orthogonal, but it is 
nearly so. Its use affords great flexibility in dealing with functions that involve many active scales: wave numbers
are grouped in dyadic blocks and averages over the dyadic blocks are performed.
The Littlewood-Paley decomposition  is defined in terms of a 
smooth partition of unity in Fourier space. This partition is constructed starting from a nonnegative, nonincreasing, radially symmetric function $\phi_{(0)} ({\mathbf k}) = \phi_{(0)}(k)$, that equals $1$ for $k \le \frac{5}{8}k_0$ and vanishes for $k\ge \frac{3}{4}k_0$. The positive number $k_0$ is just a reference wave number that fixes units.
The argument of this template 
function is then dilated,  setting $\phi_{(m)}(k) = \phi(2^{-m}k)$ and then
the template is differenced, setting  $\psi_{(0)}(k) = \phi_{(1)}(k) - 
\phi_{(0)}(k)$, and then $\psi_{(m)}(k) =\psi_{(0)}(2^{-m}k) = \phi_{(m+1)}(k) - \phi_{(m)}(k)$ for all integers $m$. The functions $\psi_{(m)}(k)$ are identically one for $k\in [\frac{3}{4}2^mk_0,  \frac{5}{4}2^mk_0]$, and vanish outside the 
interval $[\frac{5}{8}2^mk_0, \frac{3}{2}2^mk_0 ]$. The relationship   
$1 = \phi_{(m)}({\mathbf k}) + \sum_{n=m}^{\infty}\psi_{n}({\mathbf k})$ holds for any integer $m$. One defines the Littlewood-Paley operators $S^{(m)}$ and $\Delta_{n}$ as multiplication, in Fourier representation,
by $\phi_{(m)}({\mathbf k})$ and, respectively by $\psi_{(n)}({\mathbf k})$. 
Symbolically this means that the identity operator is written as $I = S^{(m)} +
 \sum_{n=m}^{\infty}\Delta_{n}$. The Littlewood-Paley decomposition of a function $F$ is $F = S^{(m)}F + \sum_{n=m}^{\infty}\Delta_{n}F$. For mean-zero
 functions that
decay at infinity, the terms $S^{(m)}$ becomes negligible when $m\to -\infty$ and therefore, for such functions one can write $F = \sum_{n= -\infty}^{\infty}\Delta_{n}F$. It is easy to see that for each fixed $k>0$ at most three 
$\Delta_{n}$ do not vanish in their Fourier representation at $k$ (i.e. the conditions $k\in [\frac{5}{8}2^nk_0, \frac{3}{2}2^nk_0 ]$ can be satisfied by at 
most three integers $n$, because $n\in [-1 + \log_2(\frac{k}{k_0}), 1 +\log_2(\frac{k}{k_0})]$). The operators $S^{(m)}$ and $\Delta_{n}$ can be 
viewed as convolution operators. In particular, for every $n = \pm 1, \pm2, \dots $, $\Delta_{n} = \int d{\mathbf h}\Psi_{(n)}({\mathbf h})\delta_{\mathbf h}$. Here $\Psi_{(n)}$ is the function whose Fourier transform is $\psi_{(n)}$,
$\widehat{\Psi_{(n)}} = \psi_{(n)}$, and $\delta_{\mathbf h}$ is the finite 
difference operator, $(\delta_{{\mathbf h}}F)({\mathbf r}) = F({\mathbf r}-{\mathbf h}) - F({\mathbf r})$. Thus $\Delta_n$ is a weighted sum of finite 
difference operators at scale $2^{-n}k_0^{-1}$ in physical space, ($k_0^{-1}$
provides thus an (arbitrary) length unit).

The Littlewood-Paley decomposition of the solutions of the quasi-geostrophic
equation is performed at each instance of time, 
$$
\theta ({\mathbf r},t) = \sum_{n=-{\infty}}^{\infty}\theta_{(n)}({\mathbf r},t). 
$$
We wrote for ease of notation $\theta_{(n)}$ instead of $\Delta_n\theta$, thus,
$$
\theta_{(n)}({\mathbf{r}},t) = \int d{\mathbf h}\Psi_{(n)}({\mathbf h})\delta_{\mathbf h}(\theta)({\mathbf r},t). 
$$
The finite difference is taken at equal times, $\delta_{\mathbf h}(\theta)({\mathbf r},t) = \theta({\mathbf r} -{\mathbf h}, t)-\theta ({\mathbf r},t)$. There is an analogous decomposition for the velocity ${\mathbf v}$ and the forcing term $f$. In particular, in Fourier 
variables, the Littlewood-Paley components of ${\mathbf v}$ are given by 
$\widehat{{\mathbf v}_{(n)}}({\mathbf k},t) = {\psi}_{(n)}({\mathbf k})\widehat{\mathbf v}({\mathbf k},t)$. The Littlewood-Paley spectrum \cite{LP} is
\begin{equation}
E_{LP}(k) = \frac{1}{k}\sum_{-1 + \log_2(\frac{k}{k_0})\le n \le 1 +\log_2(\frac{k}{k_0})}\langle | \widehat{{\mathbf v_{(n)}}}({\mathbf k}, t)|^2\rangle \label{elp}
\end{equation}
where $\langle \dots\rangle$ is 
space-time average.  The relation to the usual energy spectrum is 
straighforward. Because
\begin{equation}
{\widehat {\mathbf v}}({\mathbf k}, t) = \sum_{-1 + \log_2(\frac{k}{k_0})\le n \le 1 +
\log_2(\frac{k}{k_0})}{\widehat{\mathbf v}_{(n)}}({\mathbf k},t),\label{dec}
\end{equation}
it follows that the usual energy spectrum
\begin{equation}
E(k) = \frac{1}{k}\langle | \widehat{{\mathbf v}}({\mathbf k}, t)|^2\rangle \label{esp}
\end{equation}
satisfies
\begin{equation}
E(k)\le 3E_{LP}(k).\label{ineqsp}
\end{equation}
Clearly, because the functions ${\psi}_{(n)}$ are non-negative and bounded by $1$, one also has
\begin{equation}
E(k) \ge \frac{1}{3} E_{LP}(k).\label{lineqesp}
\end{equation}

The temporal evolution of the system induces an evolution
\begin{equation}
\partial_t \theta_{(n)} + {\mathbf v}\cdot\nabla \theta_{(n)} + 
w_E\Lambda \theta_{(n)}  = R_{(n)}\label{eqlp}
\end{equation}
where 
\begin{equation}
R_{(n)} = f_{(n)} + \int d{\mathbf h}\Psi_{(n)}({\mathbf h}){\mathbf \nabla}_{\mathbf h}\cdot(\delta_{\mathbf h}({\mathbf v})\delta_{\mathbf h}\theta )\label{rn}
\end{equation}
and $f_{(n)}$ is the Littlewood-Paley component of the forcing term.   Multiplying
(\ref{eqlp}) with $\theta_{(n)}$ and taking space-time average, one obtains
the balance
\begin{equation}
w_E k \langle\left | \widehat{\theta_{(n)}}({\mathbf k},t)\right |^2\rangle 
= \langle R_{(n)}({\mathbf r},t) \theta_{(n)}({\mathbf r},t)\rangle .\label{bal}
\end{equation}   
Let us consider the case when the forcing term has
a limited support in Fourier space, $\widehat{f}({\mathbf k},t) = 0$
for $k\notin [k_a, k_b]$, with $0< k_a< k_b <\infty$. The inverse cascade region will be described by wave 
numbers smaller than the minimal injection wave number $k_a$. The inverse cascade region corresponds thus, in the Littlewood-Paley decomposition, to indices $n>-\infty$ that satisfy  $2^{n+1}k_0<k_a$. We show now that the right hand side of the equation (\ref{bal}) is 
bounded above, uniformly for all such  $n> -\infty$. Because we are in a 
region where $f_{(n)} =0$, the term in the right-hand side of (\ref{bal}) can 
be written, after one integration by parts, as 
$$
\langle R_{(n)}({\mathbf r},t) \theta_{(n)}({\mathbf r},t)\rangle =
$$
\begin{equation}
 - \int d{\mathbf h}\nabla_{\mathbf h}\Psi_{(n)}({\mathbf h})\langle \delta_{\mathbf h}({\mathbf v})({\mathbf r},t) \delta_{\mathbf h}({\theta })({\mathbf r},t)\theta_{(n)}({\mathbf r},t)\rangle .\label{rntheta}
\end{equation} 
This is a weighted sum of triple correlations. We will analyze each of 
the three terms involved in it, in an elementary but rigorous fashion. 
Because we aim at an upper bound, we will not try to optimize the prefactors.
Using the Fourier inversion formula
$$
\theta_{(n)}({\mathbf r},t) = (2\pi)^{-2}\int d{\mathbf{k}}e^{i{{\mathbf {r}}
\cdot{\mathbf {k}}}}\psi_{(n)}({\mathbf k}){\widehat{\theta}}({\mathbf k}, t),
$$
the term $\theta_{(n)}$ is bounded pointwise by applying the Schwartz 
inequality: 
\begin{equation}
|\theta_{(n)}({\mathbf r},t)|\le (2\pi)^{-2}\|\psi_{(n)}\|\|\widehat{\theta}\|,
\label{thetab}
\end{equation} 
with $\|\cdots\|$
the mean square norm. Using the fact that $\psi_{(n)}$ is a dilate of 
$\psi_{(0)}$, we get 
\begin{equation}
|\theta_{(n)}({\mathbf r},t)|\le c_{\psi}2^n E(t)^{\frac{1}{2}},\label{thetan} 
\end{equation}
where $c_{\psi}^2 = (2\pi)^{-2}\int d{\mathbf k}|\psi_{(0)}({\mathbf k})|^2$ and $E(t) = \int d{\mathbf r}|\theta ({\mathbf r},t)|^2 = \int d{\mathbf r} |{\mathbf v}({\mathbf r},t)|^2$ is the instaneous total energy.
In the equality above we used Plancherel's identity $\|F\| = (2\pi)^{-1}\|\widehat{F}\|$, and (\ref{tveq}). In order to 
bound the other two terms we note that, from Plancherel, we have 
$$
\int d{\mathbf r} \left |\delta_{\mathbf h} \theta ({\mathbf r},t)\right |^2 =(2\pi)^{-2}\int d{\mathbf k} \left |e^{-i{\mathbf h}\cdot{\mathbf k}} - 1|^2\right |{\widehat \theta}({\mathbf k},t)|^2. 
$$
Using $|e^{-i{\mathbf h}\cdot{\mathbf k}} - 1|^2\le 4 hk$, we deduce 
\begin{equation}
\int d{\mathbf r} \left |\delta_{\mathbf h} \theta ({\mathbf r},t)\right |^2 \le 4h\eta (t),\label{deltatheta}
\end{equation} 
where $\eta(t) = \int d{\mathbf r}\theta({\mathbf r},t)\Lambda\theta({\mathbf r},t)$. The term involving $\delta_{\mathbf h}{\mathbf v}$ is bounded using the same argument. In view of (\ref{tveq}), the bound is by the same quantity:
\begin{equation}
\int d{\mathbf r} \left |\delta_{\mathbf h}{\mathbf {v}} ({\mathbf r},t)\right |^2 \le 4h\eta (t),\label{deltav}
\end{equation} 
Putting the three inequalities (\ref{thetan}, \ref{deltatheta}, \ref{deltav}) 
together with the Schwartz inequality, we deduce that the triple correlation 
term that is integrated in (\ref{rntheta}) obeys 
\begin{equation}
\left |\langle \delta_{\mathbf h}({\mathbf {v}})({\mathbf {r}},t)\delta_{\mathbf {h}}(\theta)({\mathbf {r}},t)\theta_{(n)}({\mathbf {r}},t)\rangle\right|
\le 4c_{\psi}2^n h\langle E(t)^{\frac{1}{2}}\eta (t)\rangle\label{inter}
\end{equation}
In view of the fact that the functions $\Psi_{(n)}$ are dilates of a fixed function, we deduce that 
\begin{equation}
|\langle R_{(n)}({\mathbf {r}},t)\theta_{(n)}({\mathbf {r}},t)\rangle | \le 2^{n+ 1}C_{\psi} \eta E^{\frac{1}{2}}. \label{rnbo}
\end{equation} 
Here  $E$ is the maximum total (not per unit volume) kinetic energy on the 
time interval, $E = \sup_t E(t)$. The constant 
\begin{equation}
C_{\psi} = 2c_{\psi}\int d{\mathbf {h}}h|\nabla_{\mathbf {h}}\Psi_{(0)}| = c_0k_0
\label{cpsi}
\end{equation} 
is proportional to $k_0$ and depends on the choice of the Littlewood-Paley template $\psi_{(0)}$ only through the non-dimensional positive absolute 
constant $c_0$. 
The number 
$$
\eta = \langle \eta(t)\rangle
$$ 
is related to the long time dissipation. It can be bound in terms of the forcing term using the total balance 
$$
\frac{1}{2}\frac {d}{dt} E(t) + w_E\eta (t) = \int d{\mathbf r}f({\mathbf r},t)\theta (r,t),
$$ 
which follows from (\ref{qg}) after multiplication by $\theta$ and integration.Writing the integral in Fourier variables and using the fact that the support of the forcing is bounded below by $k_a>0$ one obtains the bound 
\begin{equation}
\eta \le w_E^{-2}k_a^{-1}\langle |f({\mathbf r},t)|^2\rangle \label{etab}.
\end{equation} 
This bound diverges for very large scale 
forcing, i.e. when $k_a\to 0$. 
Nevertheless, because of the presence of the coefficient $2^{n+1}$ in (\ref{rnbo}) and the fact that $2^{n+1}k_0\le k_a$ in the inverse cascade region, the total bound on the spectrum does not diverge as $k_a\to 0$: inserting (\ref{etab})
in (\ref{rnbo}) and using (\ref{cpsi}) we get
\begin{equation}
|\langle R_{(n)}({\mathbf {r}},t)\theta_{(n)}({\mathbf {r}},t)\rangle | \le c_0w_E^{-2}E^{\frac{1}{2}}\langle |f({\mathbf r},t)|^2\rangle. \label{rnb}
\end{equation} 
Now, using (\ref{rnb}) in (\ref{bal}) and recalling the definition (\ref{elp})
and the inequality (\ref{ineqsp}) we obtain 
\begin{equation}
E(k) \le Ck^{-2}\label{bound}
\end{equation}
for all $k < k_a$. This is the main result of this letter. The constant
has units of length per time squared and is given by
\begin{equation}
C= 3c_0E^{\frac{1}{2}}w_E^{-3}\langle |f({\mathbf r},t)|^2\rangle. \label{c}
\end{equation}

The upper bound proved in this letter holds in greater 
generality than presented here. First of all, the spectrum of the forces need not be confined to the band $[k_a, k_b]$. The role played by $(k_a)^{-1}$ is then played by the ratio $\langle {k}^{-1}|{\widehat{f}}({\mathbf {k}},t)|^2\rangle\{\langle   |{\widehat {f}}({\mathbf {k}},t)|^2\rangle\}^{-1}$. Secondly, the results and methods apply to a much wider class of quasi-geostrophic equations. In fact, the quasi-geostrophic equation chosen here is the simplest version
of a class of equations in which the potential vorticity $q$ is advected by a 
a three dimensional velocity field that can be derived from a stream function
$\psi$ (not to be confused with our Littlewood-Paley cutoff functions). The 
velocity has no vertical component ${\mathbf v} = (u,v) = (-\partial_y \psi,
\partial_x \psi)$. The potential vorticity $q$ is advected
following horizontal trajectories, $\partial_t q + {\mathbf v}\cdot\nabla q +
\beta v = F$, where $F$ includes sources and damping. The potential vorticity 
and the stream function are functions of three space variables $(x,y,z)$; 
the relation $q = (\partial_{xx}+\partial_{yy}+ \partial_{zz})\psi$
closes this equation. The boundary conditions at $z=0$ are (\ref{qg})
with $\theta = \partial_z\psi$. One decomposes
$\psi = \psi_B + \psi_N$ in a sum of a harmonic function $\psi_B$,
$(\partial_{xx}+\partial_{yy}+ \partial_{zz})\psi_B = 0$, $\partial_z {\psi_{B}}_{|\, z= 0} = \theta$ and
a function that satisfies homogeneous Neumann boundary conditions, 
$(\partial_{xx}+\partial_{yy}+ \partial_{zz})\psi_N = q$, $\partial_z {\psi_{N}}_{|\, z = 0} = 0$. The ensuing equations on the boundary can be analyzed 
as above, using properties of the smooth evolution of $q$. 

Two main ingredients  
were used in the proof. The first one is the way in which the 
spectrum of the potential temperature is related to the energy spectrum.
The second, and the essential ingredient is the fact that the 
relaxation time at wave number $k$ is roughly $(w_E k)^{-1}$ in the range of 
wave numbers considered. This dependence is an important by-product of the 
quasi-geostrophic model. In contradistinction with  direct cascade models 
where there is a dissipation anomaly, in the quasi-geostrophic models the 
coefficient $w_E$ is not vanishingly small. Moreover, the physical forcing 
amplitude obtained from an Ekman boundary layer is proportional to $w_E$.
The explicit presence of the large scale 
term $E^{\frac{1}{2}}$ in the prefactor $C$ is a reflection of the fact that
the $k^{-2}$ spectrum is modified near $k = 0$. The same is true for the 
Kolmogorov-Kraichnan spectrum: the integrals diverge and require an infrared
cutoff. Such a cutoff can be achieved mathematically in  
two different ways. One may impose a smallest wave number $k_{min}$ and 
boundary conditions; or one can modifiy  the dissipation law so that, in the 
limit $k\to 0$ one has a finite relaxation time. In either case
one can prove an {\it a priori} upper bound on $E$ depending on the 
forcing and dissipation mechanism. The fact that the large scales are nearly conservative, with finite
energy, was used in \cite{Sw} to compare the $k^{-2}$ and $k^{-\frac{5}{3}}$
spectra with the same largest scale energy. Our upper 
bound (\ref{bound}) confirms theoretically the separation of the
two spectra when the injection length scales is small enough. This is 
indeed the case in the experiment \cite{Sw}: forcing was applied through 
120 holes of diameters of .25 cm,  some 8 times larger than the 
Ekman boundary layer length. 

In summary, we have proved that the energy spectrum of a forced 
surface quasi-geostrophic equation is bounded above by $Ck^{-2}$ 
for wave numbers that are smaller than the force's injection wave number.
Such a bound distinguishes the quasi-geostrophic model from a two-dimensional
Navier-Stokes model, and agrees with the recent experimental evidence of 
\cite{Sw}.
 
The author acknowledges useful discussions with R. T. Pierrehumbert and
with H.L. Swinney. This research is supported in part by NSF DMS-0202531
and by the ASCI Flash Center at the University of Chicago under DOE
contract B341495.

\end{document}